\documentclass[conference]{IEEEtran}
\IEEEoverridecommandlockouts
\usepackage{cite}
\usepackage{amsmath,amssymb,amsfonts}
\usepackage{graphicx}
\usepackage{textcomp}
\usepackage{xcolor}
\usepackage{booktabs}                        
\usepackage{bm}                              
\usepackage{color}
\usepackage[colorlinks,linkcolor=blue]{hyperref}
\usepackage{enumerate}                       
\usepackage{epstopdf}                        
\usepackage{mathtools}                       
\usepackage{mathrsfs}
\usepackage{verbatim} 
\usepackage{url}
\usepackage[ruled,norelsize]{algorithm2e} 
\ifCLASSOPTIONcompsoc            
  \usepackage[caption=false,font=normalsize,labelfont=sf,textfont=sf]{subfig}
\else
  \usepackage[caption=false,font=footnotesize]{subfig}
\fi
\def\BibTeX{{\rm B\kern-.05em{\sc i\kern-.025em b}\kern-.08em
    T\kern-.1667em\lower.7ex\hbox{E}\kern-.125emX}}

\makeatletter
\newcommand{\removelatexerror}{\let\@latex@error\@gobble}
\makeatother

\begin{document}

\title{Multi-target Detection for Reconfigurable Holographic Surfaces Enabled Radar
}

\author{\IEEEauthorblockN{Xiaoyu~Zhang$^{*}$,
Haobo~Zhang$^{*}$, Ruoqi~Deng$^{*}$, Liang~Liu$^{\dag}$, and
Boya~Di$^{*}$}\\
\IEEEauthorblockA{$^{*}$State Key Laboratory of Advanced Optical Communication Systems and Networks, \\School of Electronics, Peking University, Beijing, China.\\
$^{\dag}$The Hong Kong Polytechnic University, Hong Kong SAR, China.\\
Email: $^{*}$\{xiaoy.zhang, haobo.zhang, ruoqi.deng, diboya\}@pku.edu.cn,
$^{\dag}$liang-eie.liu@polyu.edu.hk}
\thanks{This work was supported in part by the National Key R\&D Project of China
under Grant No. 2022YFB2902800, National Science Foundation under Grant
62271012 and 62227809, and Beijing Natural Science Foundation under Grant
4222005 and L212027.}
}



\maketitle

\begin{abstract}
	Multi-target detection is one of the primary tasks in radar-based localization and sensing, typically built on phased array antennas. 
	However, the bulky hardware in the phased array restricts its potential for enhancing detection accuracy, since the cost and power of the phased array can become unaffordable as its physical aperture scales up to pursue higher beam shaping capabilities.
	To resolve this issue, we propose a radar system enabled by reconfigurable holographic surfaces~(RHSs), a novel meta-surface antenna composed of meta-material elements with cost-effective and power-efficient hardware, which performs multi-target detection in an adaptive manner. 
	Different from the phase-control structure in the phased array, the RHS is able to apply beamforming by controlling the radiation amplitudes of its elements. 
	Consequently, traditional beamforming schemes designed for phased arrays cannot be directly applied to RHSs due to this structural difference. 
	To tackle this challenge, a waveform and amplitude optimization algorithm (WAOA) is designed to jointly optimize the radar waveform and RHS amplitudes in order to improve the detection accuracy. 
	Simulation results reveal that the proposed RHS-enabled radar increases the probability of detection by 0.13 compared to phased array radars when six iterations of adaptive detection are performed given the same hardware cost. 
\end{abstract}


\section{Introduction}
High-accuracy localization and sensing are essential in facilitating the advancement of 
indoor navigation in large buildings, inventory management in factories,
and other applications within the context of next-generation wireless systems~\cite{Intro_6g},~\cite{ris}.
To fulfill this vision, radars are recognized as a fundamental enabler,
which conventionally rely on phased arrays antennas.
The phased array, comprised of an array of antenna elements, empowers independent phase control of the fed signal at each element through a phase shifter, thus enabling electrically controlled beam steering for the sensing of targets in different directions.

However, the intrinsic characteristics of the phased array have limited its development in future high-demanding sensing applications. 
Specifically, on the one hand, the antenna gain that is positively correlated to the sensing accuracy is inherently constrained by the physical aperture of the antenna. 
On the other hand, 
in light of the use of costly and power-consuming hardware components like phase shifters, scaling up the physical aperture of the phased array leads to an unaffordable increase in manufacturing expenses and power consumption.
Thus, there is a pressing need for novel antennas with low cost and power consumption and corresponding signal processing techniques. 

Recently, the advancement of reconfigurable holographic surfaces~(RHSs) brings a promising solution to the above issues~\cite{RHS_intro}. The RHS is a type of meta-material antenna that can form a desired beampattern by independently adjusting the amplitude of the radiated signal at each element. 
Different from the phase control in the phased array, the amplitude control at each RHS element is realized by a simple diode-based circuit, which is much more cost-effective and power-efficient.
Hence, the RHS can be a favorable candidate for the next-generation radar antenna to support high-accuracy localization and sensing.

Existing works have already explored the use of RHS in radar sensing applications~\cite{RHS_sens_1},~\cite{RHS_sens_2}. In~\cite{RHS_sens_1}, a microwave camera is synthesized using the RHS for imaging task, where the RHS is designed to form a beam pattern towards a specific direction at each time in order to scan the scene. In~\cite{RHS_sens_2}, an RHS-enabled radar is developed for target detection by maximizing the signal-to-noise-ratio (SNR) of the reflected signal at a certain direction. 
However, these works only focus on the RHS-aided detection of a single target at a time, research on the simultaneous detection of multiple targets is still lacking. 


In this paper, we investigate multi-target detection enabled by the RHSs, where two RHSs are closely deployed to serve as transmit and receive antennas, respectively. 
By carefully designing the amplitudes of the two RHSs, the proposed system is able to simultaneously detect multiple targets with high accuracy. 
However, this task is non-trivial due to the following two reasons. 
First, due to the difference in hardware structure between the RHSs and phased arrays, traditional beamforming schemes designed for phased arrays with phase control cannot be directly applied to RHSs with amplitude control.
Second, it is challenging to jointly optimize the radar waveform and RHS amplitudes considering the coupling between them in terms of the detection performance.

To confront these challenges, we formulate a relative entropy based optimization problem to jointly optimize the radar waveform and the RHS amplitudes, and design a waveform and amplitude optimization algorithm (WAOA) to solve this problem. Simulation results are presented to verify the effectiveness of the proposed algorithm. Moreover, it is demonstrated that the proposed RHS-enabled scheme can enhance the detection performance for multiple targets compared to traditional phased array based scheme.

\section{System Model}
\label{s_system}
\subsection{Scenario Description}
\begin{figure}[!t]
	\setlength{\abovecaptionskip}{-3pt}
	\setlength{\belowcaptionskip}{-25pt}
	\centering
	\includegraphics[width=2.4in]{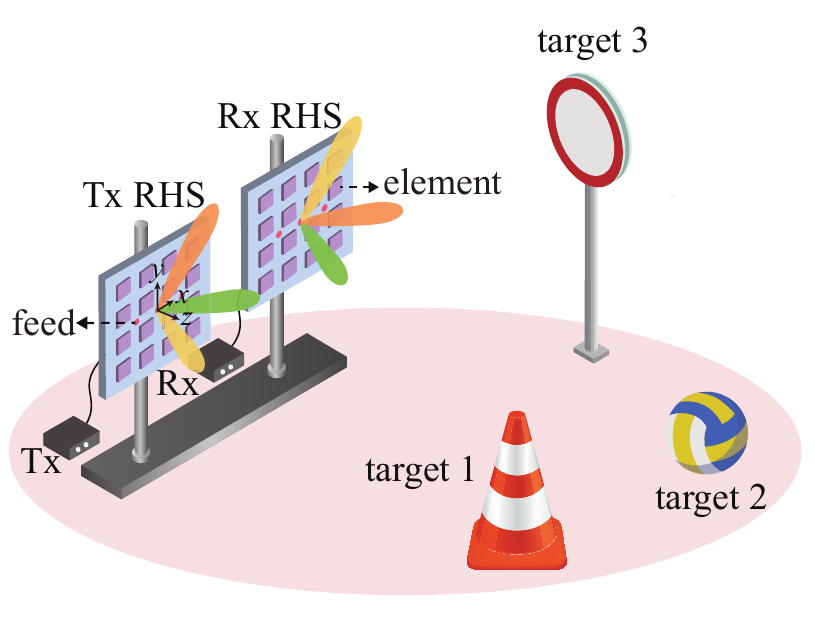}
	\caption{System model for the RHS-enabled radar system.}
	\label{fig_sys_mod}
	\vspace{-6.5mm}
\end{figure}

Consider an RHS-enabled radar system for multi-target detection as shown in Fig. \ref{fig_sys_mod}. The system is composed of a transmitter (Tx), a receiver (Rx), an RHS connected to the Tx (Tx RHS), and another RHS connected to the Rx (Rx RHS). The two RHSs are closely deployed, 
and the targets to be detected are in the far-field of the radar system.

The aim of the proposed radar system is to detect all the targets without the prior knowledge of their number and locations. 
The detection process consists of multiple cycles, and each cycle consists of the following three steps~\cite{multi-targ_detec_1}:
\begin{itemize}
	\item \textbf{Optimization:} The radar waveform and RHS amplitudes are optimized based on the prior information obtained from previous cycles. In the first cycle, these variables are initialized randomly. The details of the optimization problem will be presented in Section~\ref{prob_form} and \ref{algo_desi}.
	\item \textbf{Transmission and reception:} The radar signals are transmitted and received in this step, which will be discussed in Subsections \ref{RHS_modl} and \ref{recv_modl}.
	\item \textbf{Determination:} In this step, we update the estimation of the target information by analyzing received signals. 
	The termination condition is then checked to determine whether to end the detection process or move on to the next cycle, which will be introduced in Subsection~\ref{dete_modl}.
\end{itemize}




\subsection{Model of Reconfigurable Holographic Surfaces}
\label{RHS_modl}
The RHS is a special kind of leaky wave antenna composed of feeds and an array of meta-material elements, as shown in Fig. \ref{fig_sys_mod}.
The RHS operates through serial feeding.
Specifically, when the RHS serves as a transmit antenna, the radar signals are injected via the feeds, and propagate along the RHS elements to excite the elements one by one. 
When an element is excited, the signal is radiated as leaky wave at this element.
And the amplitude of the radiated wave at each element can be independently designed between $[0,1]$ to steer radiation beam. 

Consider a two-dimensional (2D) Tx RHS with $L_t$ feeds and $N_t$ 
elements. The transmitted signal $\boldsymbol{S}^{(c)}\in\mathbb{C}^{N_t\times I_t}$ at the $c$-th cycle can be formulated as
\begin{equation}\label{tx_mat_model}
	\boldsymbol{S}^{(c)} = \boldsymbol{\Psi}^{t(c)} (\boldsymbol{Q}^t \circ \boldsymbol{\Gamma}^t) \boldsymbol{X}^{(c)},
\end{equation}
where $\boldsymbol{X}^{(c)}\in\mathbb{C}^{L_t\times I_t}$ denotes the radar signal at the $c$-th cycle with $I_t$ being the snapshot number, $\boldsymbol{\Psi}^{t(c)} = diag\left\{ \boldsymbol{\psi}^{t(c)} \right\}$ with $\boldsymbol{\psi}^{t(c)} = [\psi_1^{t(c)},\cdots,\psi_{N_t}^{t(c)}]^T$ being the amplitude vector of the Tx RHS. $\boldsymbol{Q}^t=\{ q_{n,l}^t \} \in \mathbb{C}^{N_t\times L_t}$ and $\boldsymbol{\Gamma}^t=\{ \gamma_{n,l}^t \} \in \mathbb{C}^{N_t\times L_t}$ are the phase shift matrix and the attenuation matrix, where $q_{n,l}^t=e^{-j2\pi\nu D_{n,l}/\lambda}$ and $\gamma_{n,l}^t=e^{-\alpha D_{n,l}}$ respectively denote inherent phase shift and amplitude attenuation of radar signal when propagating from the $l$-th feed to the $n$-th element with distance $D_{n,l}$, refractive index $\nu$, and attenuation factor $\alpha$.

Due to the reciprocity of antennas \cite{RHS_sens_2}, the received signal of a 2D Rx RHS with $L_r$ feeds and $N_r$ 
elements given the reflected signal from the targets $\boldsymbol{V}^{(c)}\in\mathbb{C}^{N_r\times I_r}$ can be expressed as
\begin{equation}\label{rx_mat_model}
	\boldsymbol{Y}^{(c)} = \left[\boldsymbol{\Psi}^{r(c)}(\boldsymbol{Q}^r \circ \boldsymbol{\Gamma}^r)\right]^T \boldsymbol{V}^{(c)},
\end{equation}
where $\boldsymbol{\Psi}^{r(c)} = diag\left\{ \boldsymbol{\psi}^{r(c)} \right\}$ being the amplitude vector at the $c$-th cycle with $\boldsymbol{\psi}^{r(c)} = [\psi_1^{r(c)},\cdots,\psi_{N_r}^{r(c)}]^T$, $\boldsymbol{Q}^r=\{ q_{n,l}^r \} \in \mathbb{C}^{N_r\times L_r}$ denotes the phase shift matrix, $\boldsymbol{\Gamma}^r =\{ \gamma_{n,l}^r \} \in \mathbb{C}^{N_r\times L_r}$ is the attenuation matrix, and $I_r$ is the snapshot number of the received signal vector.








\subsection{Received Signal Model}\label{recv_modl}
Assume the target number in the interesting area is $K$, which is unknown. And we assume that the multiple targets are located in different angular grids and have the same range.
The reflected signal at the $c$-th cycle can be given by
\begin{equation}\label{rx_sig_model}
	\boldsymbol{V}^{(c)} = \sum_{k=1}^{K} \beta_{k} \boldsymbol{a}_r(\theta_{k},\varphi_{k})\boldsymbol{a}_t^T(\theta_{k},\varphi_{k}) \boldsymbol{S}^{(c)} + \boldsymbol{W}^{(c)},
\end{equation}
where $\beta_{k}$ is the reflection coefficient of the $k$-th target, $\boldsymbol{a}_t(\theta_{k},\varphi_{k})\in\mathbb{C}^{N_t\times 1}$ and $\boldsymbol{a}_r(\theta_{k},\varphi_{k})\in\mathbb{C}^{N_r\times 1}$ are steering vectors with $\theta_{k}$ and $\varphi_{k}$ being the azimuthal and polar angles, respectively. $\boldsymbol{W}^{(c)}\in\mathbb{C}^{N_r\times I_r}$ denotes the received noise, whose row vectors are independent and identically distributed circularly symmetric complex Gaussian random vectors with zero mean and covariance matrix $\sigma^2\boldsymbol{I}^{I_r\times I_r}$.

The received signal in vector form $\boldsymbol{y}^{(c)} = \text{vec}\left\{\boldsymbol{Y}^{(c)}\right\}$ at the $c$-th cycle can be given by
\begin{equation}
	\begin{aligned}
		\boldsymbol{y}^{(c)} = \boldsymbol{u}^{(c)}(K,\boldsymbol{x}^{(c)},\boldsymbol{\psi}^{t(c)},\boldsymbol{\psi}^{r(c)}) + \boldsymbol{w}^{(c)},
	\end{aligned}
\end{equation}
where $\boldsymbol{u}^{(c)}(K,\boldsymbol{x}^{(c)},\boldsymbol{\psi}^{t(c)},\boldsymbol{\psi}^{r(c)}) = \sum_{k=1}^{K} \beta_{k} \text{vec}\{ [\boldsymbol{\Psi}^{r(c)}(\boldsymbol{Q}^r \circ \boldsymbol{\Gamma}^r)]^T \boldsymbol{a}_r(\theta_{k},\varphi_{k}) \boldsymbol{a}_t^T(\theta_{k},\varphi_{k}) \boldsymbol{\Psi}^{t(c)} (\boldsymbol{Q}^t \circ \boldsymbol{\Gamma}^t) \boldsymbol{X}^{(c)} \}$ represents the effective signal term with $\boldsymbol{x}^{(c)} = \text{vec}\big\{ \boldsymbol{X}^{(c)} \big\}$, and $\boldsymbol{w}^{(c)}=\text{vec}\big\{ \left[\boldsymbol{\Psi}^r(\boldsymbol{Q}^r \circ \boldsymbol{\Gamma}^r)\right]^T \boldsymbol{W}^{(c)} \big\}$ is the noise term.



\subsection{Detection Model}\label{dete_modl}
The proposed radar system performs multi-target detection by using multi-hypothesis testing. Specifically, we discretize the interesting area into $J$ angular grids, namely $\left\{(\theta_j,\varphi_j), j=1,\cdots,J\right\}$, and formulate a total number of $N_H$ hypotheses $\mathcal{H}_0,\mathcal{H}_1,\cdots,\mathcal{H}_{N_H-1}$. $\mathcal{H}_0$ represents the null hypothesis. $\mathcal{H}_j~(j\neq 0)$ can be described by $\boldsymbol{v}(\mathcal{H}_j)=\left[v_1(\mathcal{H}_j),\cdots,v_{\mathcal{I}(\mathcal{H}_j)}(\mathcal{H}_j)\right]$, meaning that there are $\mathcal{I}(\mathcal{H}_j)$ targets and the $i$-th target is located in the $v_i(\mathcal{H}_j)$-th grid. 
In each cycle of the detection process, we update the probabilities of the hypotheses. 
The process ends when the probability of a certain hypothesis exceeds a specific threshold while probabilities of other hypotheses are below a lower limit.
The hypothesis with the highest probability is accepted as the final detection result.

First, we define the prior probability of hypothesis $\mathcal{H}_j$ in the first cycle as
\begin{equation}
	p^{(1)}(\mathcal{H}_j) = \frac{1}{(K_m+1) N_{\mathcal{I}(\mathcal{H}_j)}},
\end{equation}
where $K_m$ is the maximum number of potential targets, and $N_{\mathcal{I}(\mathcal{H}_j)}$ is the number of hypotheses assuming $\mathcal{I}(\mathcal{H}_j)$ targets exist.
Given the received signal $\boldsymbol{y}^{(c)}$ at the $c$-th cycle,
the prior probability at the $(c+1)$-th cycle can be expressed as
\begin{equation}\label{post_p}
	p^{(c+1)}(\mathcal{H}_j) = \frac{p^{(c)}(\mathcal{H}_j) p(\boldsymbol{y}^{(c)} | \mathcal{H}_j)}{\sum_i p^{(c)}(\mathcal{H}_i) p(\boldsymbol{y}^{(c)} | \mathcal{H}_i)}. 
\end{equation}

Denote the effective signal under hypothesis $\mathcal{H}_j$ at the $c$-th cycle as $\boldsymbol{u}_j^{(c)} = \boldsymbol{u}^{(c)}(\mathcal{I}(\mathcal{H}_j),\boldsymbol{x}^{(c)},\boldsymbol{\psi}^{t(c)},\boldsymbol{\psi}^{r(c)})$.
The likelihood function $p(\boldsymbol{y}^{(c)} | \mathcal{H}_j)$ can be defined as
\begin{equation}
	\begin{aligned}
		p(\boldsymbol{y}^{(c)} | \mathcal{H}_j) =& \frac{1}{\sqrt{(2\pi)^{L_rI_r} \text{det}\left(\boldsymbol{\Sigma}\right)}} \cdot \\
		& \exp \left\{ -\frac{1}{2} \left[ \boldsymbol{y}^{(c)}-\boldsymbol{u}_j^{(c)} \right]^H \boldsymbol{\Sigma}^{-1} \left[ \boldsymbol{y}^{(c)}-\boldsymbol{u}_j^{(c)} \right] \right\}.
	\end{aligned}
\end{equation}
In the above equation, $\boldsymbol{\Sigma} = \boldsymbol{I}_{I_r\times I_r} \otimes \boldsymbol{F}$ denotes the noise variance matrix of $\boldsymbol{w}$ with $\boldsymbol{F}=diag\big\{ \sum_{n=1}^{N_r} |\psi_n^r q_{n,l}^r \gamma_{n,l}^r|^2\sigma^2,~l=1,2,\cdots,L_r \big\}$, $\text{det}\left(\cdot\right)$ denotes the determinant operator, and $(\cdot)^{-1}$ is the inverse operator. 

\section{Problem Formulation}\label{prob_form}


\subsection{Problem Formulation}
We aim to improve detection accuracy by jointly optimizing the radar waveform $\boldsymbol{x}$ and amplitude vectors of RHSs $\boldsymbol{\psi}^t, \boldsymbol{\psi}^r$. According to~\cite{multi-targ_detec_1}, we adopt the weighted sum of relative entropy as the optimization objective. 
Specifically, the relative entropy $D(p(x)\Vert q(x))=\int_{-\infty}^{+\infty} p(x) \log \left( \frac{p(x)}{q(x)} \right) \text{d}x$ measures the distance between probability distributions $p(x)$ and $q(x)$. The larger the relative entropy, the easier the distinguishment of different hypotheses.
Therefore, we define the objective function at $(c+1)$-th cycle as the weighted sum of the relative entropy for any two hypotheses, which can be given by  
\begin{equation}\label{d_ave}
	\begin{aligned}
		\bar{d}^{(c+1)} &= \sum_{i=0}^{N_H} \sum_{j=0}^{N_H} \omega_{i,j} D\left(p(\boldsymbol{y}^{(c)} | \mathcal{H}_i) \Vert p(\boldsymbol{y}^{(c)} | \mathcal{H}_j)\right) \\
		&= \sum_{i=1}^{N_H} \sum_{j=0}^{i-1} \omega_{i,j} d^{(c+1)}(\mathcal{H}_i,\mathcal{H}_j|\boldsymbol{y}^{(c)}), %
	\end{aligned} 
\end{equation}
where the weights are chosen as $\omega_{i,j}=p^{(c+1)}(\mathcal{H}_i)p^{(c+1)}(\mathcal{H}_j)$ according to the idea that the distance between the hypotheses with higher probabilities should be given more priority in the maximization to further distinguish these hypotheses. 
And $d^{(c+1)}(\mathcal{H}_i,\mathcal{H}_j|\boldsymbol{y}^{(c)}) = D\left(p(\boldsymbol{y}^{(c)} | \mathcal{H}_i) \Vert p(\boldsymbol{y}^{(c)} | \mathcal{H}_j)\right) + D\left(p(\boldsymbol{y}^{(c)} | \mathcal{H}_j) \Vert p(\boldsymbol{y}^{(c)} | \mathcal{H}_i)\right)$ denotes the symmetric distance between $\mathcal{H}_i$ and $\mathcal{H}_j$, which can be further expressed as:

\newtheorem{proposition}{\bf Proposition}
\begin{proposition}\label{prop_1}
	$d^{(c+1)}(\mathcal{H}_i,\mathcal{H}_j|\boldsymbol{y}^{(c)})$ can be given by 
	\begin{equation}
		\begin{aligned}
			& d^{(c+1)}(\mathcal{H}_i,\mathcal{H}_j|\boldsymbol{y}^{(c)}) \\ 
			=& \left[\boldsymbol{u}_i^{(c+1)}-\boldsymbol{u}_j^{(c+1)}\right]^H \boldsymbol{\Sigma}^{-1} \left[\boldsymbol{u}_i^{(c+1)}-\boldsymbol{u}_j^{(c+1)}\right].\label{eq_prop1}
		\end{aligned}
	\end{equation}
\end{proposition}
\begin{IEEEproof}
	See Appendix \ref{proof_1}.
\end{IEEEproof}

Based on \eqref{eq_prop1}, with $\boldsymbol{O}^{(c+1)}=\left\{\boldsymbol{\psi}^{t(c+1)},\boldsymbol{\psi}^{r(c+1)},\boldsymbol{x}^{(c+1)}\right\}$ denoting the optimization variables, the optimization problem at the $(c+1)$-th cycle can be formulated as 
\begin{subequations}
	\begin{align}
		\text{P1:}\max_{\boldsymbol{O}^{(c+1)}}~& \sum_{i=1}^{N_H} \sum_{j=0}^{i-1} \omega_{i,j} d^{(c+1)}(\mathcal{H}_i,\mathcal{H}_j|\boldsymbol{y}^{(c)}),\label{P2_OF}\\
		s.t.~& \text{tr}\left\{ \boldsymbol{S}^{(c)} \boldsymbol{S}^{(c)H} \right\} \le P_M, \label{const_pow}\\
		& 0 \leq \psi_{n}^t,\psi_{n}^r \leq 1, \forall n \in \{1,\cdots,N_t\}, \label{psi_val}
	\end{align}
\end{subequations}
where constraint \eqref{const_pow} specifies that the transmitted power is smaller than the upper bound $P_M$, and \eqref{psi_val} is the constraint of RHS amplitudes.

\subsection{Problem Decomposition}
Problem (P1) is difficult to directly solve since the variables $\boldsymbol{\psi}^t$, $\boldsymbol{\psi}^r$ and $\boldsymbol{x}$ are coupled with each other\footnote{For simplicity, the superscript $(c+1)$ is omitted in the rest of the paper.}. To tackle (P1) efficiently, we decompose it into three sub-problems, which are respectively reformulated as follows: 
\subsubsection{Waveform Optimization Sub-problem}
Given $\boldsymbol{\psi}^t$ and $\boldsymbol{\psi}^r$, problem (P1) can be reformulated as:
\begin{subequations}
	\begin{align}
		\text{P}_{x}\text{:}\max_{\boldsymbol{x}}~& \boldsymbol{x}^H \boldsymbol{R}^x \boldsymbol{x},\\
	s.t.~& \boldsymbol{x}^H \boldsymbol{S}^x \boldsymbol{x} \leq P_M,
	\end{align}
\end{subequations}
where $\boldsymbol{S}^x = \sum_{n=1}^{N_t} \boldsymbol{B}_n^H \boldsymbol{B}_n$ and $\boldsymbol{R}^x = \sum_{i=1}^{N_H} \sum_{j=0}^{i-1} \omega_{i,j} \cdot \sum_{l=1}^{L_r} \frac{ (\boldsymbol{C}^x_l(\mathcal{H}_i) - \boldsymbol{C}^x_l(\mathcal{H}_j))^H (\boldsymbol{C}^x_l(\mathcal{H}_i) - \boldsymbol{C}^x_l(\mathcal{H}_j))}{(\boldsymbol{\psi}^r)^H \boldsymbol{S}^r_l \boldsymbol{\psi}^r}$ are Hermitian positive semi-definite matrices with 
\begin{equation}
	\boldsymbol{B}_n = \boldsymbol{I}_{I_t} \otimes \left[(\boldsymbol{Q}^t_n\circ\boldsymbol{\Gamma}_n)\psi_n^t\right],
\end{equation}
\begin{equation}
	\boldsymbol{S}^r_l = \boldsymbol{D}^H \boldsymbol{D}, \boldsymbol{D} = diag\left\{ q_{1,l}\gamma_{1,l},\cdots, q_{N_r,l}\gamma_{N_r,l} \right\},
\end{equation}

\begin{equation}
	\vspace{0.01in}
	\boldsymbol{C}^x_l(\mathcal{H}_i) = \sum_{k=1}^{\mathcal{I}(\mathcal{H}_i)} \beta_{k} \boldsymbol{I}_{I_r\times I_t} \otimes \left[ \boldsymbol{P}^r_{l,k} \boldsymbol{\psi}^r (\boldsymbol{\psi}^t)^T \boldsymbol{A}^{'}_k \boldsymbol{P}^t\right].
\end{equation}
Here $\boldsymbol{A}^{'}_k=diag\left\{ \boldsymbol{a}_t(\theta_{k},\varphi_{k}) \right\}$, $\boldsymbol{P}^t=\boldsymbol{Q}^t \circ\boldsymbol{\Gamma}^t$, and $\boldsymbol{P}^r_{l,k} = (\boldsymbol{Q}^r_l \circ \boldsymbol{\Gamma}^r_l \circ \boldsymbol{a}_r(\theta_{k},\varphi_{k}))^T$.

\subsubsection{Transmit Amplitude Optimization Sub-problem}
Given $\boldsymbol{x}$ and $\boldsymbol{\psi}^r$, the sub-problem can be formulated as:
\begin{subequations}
	\begin{align}
		\text{P}_{t}\text{:}\max_{\boldsymbol{\psi}^t}~& (\boldsymbol{\psi}^t)^H \boldsymbol{R}^t \boldsymbol{\psi}^t,\label{Pt_OF}\\
	s.t.~& (\boldsymbol{\psi}^t)^H \boldsymbol{S}^t \boldsymbol{\psi}^t \leq P_M,\\
	& 0 \leq \psi_{n}^t \leq 1, \forall n, \label{Pt_const1}
	\end{align}
\end{subequations}
where $\boldsymbol{S}^t = diag\left\{ ((\boldsymbol{Q}^t_n\circ\boldsymbol{\Gamma}_n)\boldsymbol{X})((\boldsymbol{Q}^t_n\circ\boldsymbol{\Gamma}_n)\boldsymbol{X})^H\right\} 
= \left[((\boldsymbol{Q}^t \circ \boldsymbol{\Gamma}^t)\boldsymbol{X})((\boldsymbol{Q}^t \circ \boldsymbol{\Gamma}^t)\boldsymbol{X})^H\right] \circ \boldsymbol{I}_{N_t}$ 
and $\boldsymbol{R}^t = \sum_{i=1}^{N_H} \sum_{j=0}^{i-1} \omega_{i,j} \sum_{l=1}^{L_r} \frac{ (\boldsymbol{C}^t_l(\mathcal{H}_i) - \boldsymbol{C}^t_l(\mathcal{H}_j))^H (\boldsymbol{C}^t_l(\mathcal{H}_i) - \boldsymbol{C}^t_l(\mathcal{H}_j))}{(\boldsymbol{\psi}^r)^H \boldsymbol{S}^r_l \boldsymbol{\psi}^r}$ are Hermitian positive semi-definite with 
\begin{equation}
		\boldsymbol{C}^t_l(\mathcal{H}_i) = \sum_{k=1}^{\mathcal{I}(\mathcal{H}_i)} \beta_{k} \boldsymbol{P}^r_{l,k} \boldsymbol{\psi}^r (\boldsymbol{P}^t\boldsymbol{X})^T \boldsymbol{A}^{'}_k.
\end{equation}

\subsubsection{Receive Amplitude Optimization Sub-problem}
Similarly, given $\boldsymbol{x}$ and $\boldsymbol{\psi}^t$, the sub-problem can be expressed as:
\begin{subequations}
	\begin{align}
		\text{P}_{r}\text{:}\max_{\boldsymbol{\psi}^r}~& \sum_{l=1}^{L_r} \frac{(\boldsymbol{\psi}^r)^H \boldsymbol{R}_l^r \boldsymbol{\psi}^r}{(\boldsymbol{\psi}^r)^H \boldsymbol{S}^r_l \boldsymbol{\psi}^r}, \label{Pr_OF}\\
		s.t.~& 0 \leq \psi_{n}^r \leq 1, \forall n,
	\end{align}
\end{subequations}
where $\boldsymbol{R}^r_l = \sum_{i=1}^{N_H} \sum_{j=0}^{i-1} \omega_{i,j} (\boldsymbol{C}^r_l(\mathcal{H}_i) - \boldsymbol{C}^r_l(\mathcal{H}_j))^H \cdot (\boldsymbol{C}^r_l(\mathcal{H}_i) - \boldsymbol{C}^r_l(\mathcal{H}_j))$ and $\boldsymbol{S}^r_l$ are Hermitian positive semi-definite which satisfy
\begin{equation}
	\boldsymbol{C}^r_l(\mathcal{H}_i) =\sum_{k=1}^{\mathcal{I}(\mathcal{H}_i)} \beta_{k} (\boldsymbol{P}^t \boldsymbol{X})^T \boldsymbol{A}^{'}_k\boldsymbol{\psi}^t \boldsymbol{P}^r_{l,k}.
\end{equation}


\section{Algorithm Design}\label{algo_desi}
In this section, a waveform and amplitude optimization algorithm (WAOA) is proposed to solve (P1) by iteratively optimizing the waveform and RHS amplitudes.
In each iteration, we successively solve sub-problems ($\text{P}_{x}$), ($\text{P}_{t}$), and ($\text{P}_{r}$). The overall algorithm is summarized in Algorithm \ref{iter_algo}. The optimization algorithms for all the sub-problems are described in the following subsections, respectively.

\begin{figure}[!t]
	\removelatexerror
	\vspace{6mm}
	\begin{algorithm}[H]
	    \caption{Waveform and Amplitude Optimization Algorithm (WAOA)}
	    \label{iter_algo}
		Set $i=0$\;
	    Randomly initialize $\boldsymbol{x}_i$, $\boldsymbol{\psi}^{t}_i$, and $\boldsymbol{\psi}^{r}_i$\;
		Calculate $\bar{d}_i$ given $\boldsymbol{x}_i$, $\boldsymbol{\psi}^{t}_i$, and $\boldsymbol{\psi}^{r}_i$ by using \eqref{d_ave}\;
	    \Repeat{$|\bar{d}_i-\bar{d}_{i-1}|\le\epsilon$}{ 
	        Set $i=i+1$\;
			Optimize $\boldsymbol{x}_i$ given $\boldsymbol{\psi}^t_{i-1}$ and $\boldsymbol{\psi}^{r}_{i-1}$\; 
			Optimize $\boldsymbol{\psi}^t_{i}$ given $\boldsymbol{x}_i$ and $\boldsymbol{\psi}^{r}_{i-1}$\; 
			Optimize $\boldsymbol{\psi}^r_{i}$ given $\boldsymbol{x}_i$ and $\boldsymbol{\psi}^t_{i}$\;
			Calculate $\bar{d}_i$ given $\boldsymbol{x}_i$, $\boldsymbol{\psi}^{t}_i$, and $\boldsymbol{\psi}^{r}_i$ by using \eqref{d_ave}\;
	    }
	\end{algorithm}
\end{figure}

\subsection{Waveform Optimization}

According to the spectral theorem, 
the Hermitian matrix $\boldsymbol{S}^x$ can be decomposed as $\boldsymbol{S}^x = \boldsymbol{U}^x\boldsymbol{D}^x(\boldsymbol{U}^x)^H$, where $\boldsymbol{D}^x$ is a diagonal matrix containing non-negative real numbers, 
and $\boldsymbol{U}^x$ is a unitary matrix.
Thus, by denoting $\boldsymbol{x}{'}=(\boldsymbol{D}^x)^{\frac{1}{2}}(\boldsymbol{U}^x)^H\boldsymbol{x}$, problem ($\text{P}_{x}$) can be reformulated as
\begin{align}
	\text{P}_{x}^{'}\text{:}\max_{\boldsymbol{x}{'}}~& (\boldsymbol{x}{'})^H \boldsymbol{R}^{x}{'} (\boldsymbol{x}{'}),\\
	s.t.~& (\boldsymbol{x}{'})^H (\boldsymbol{x}{'}) \leq P_M,
\end{align}
where $\boldsymbol{R}^{x}{'}=(\boldsymbol{D}^{x})^{-\frac{1}{2}} (\boldsymbol{U}^{x})^H \boldsymbol{R}^{x} \boldsymbol{U}^x (\boldsymbol{D}^{x})^{-\frac{1}{2}}$ is Hermitian positive semi-definite\footnote{If $\boldsymbol{S}^x$ is not positive definite, we compute the generalized inverse of $\boldsymbol{D}^x$ to approximate $(\boldsymbol{D}^x)^{-1}$.}, and $(\cdot)^{\frac{1}{2}}$ is the square root operator. 


As proved in~\cite{cf-deriv-2}, the optimal solution of problem $(\text{P}_{x}^{'})$ is 
	$\boldsymbol{x}{'}^{*} = \sqrt{P_M} ~\boldsymbol{v}^x$,
where $\boldsymbol{v}^x$ is the normalized eigenvector corresponding to the largest eigenvalue of $\boldsymbol{R}^{x}{'}$.
Therefore, the optimal solution to problem $(\text{P}_{x})$ can be given by 
\begin{equation}\label{x_star}
	\boldsymbol{x}^{*} = \left[(\boldsymbol{D}^x)^{\frac{1}{2}}(\boldsymbol{U}^x)^H\right]^{-1} \boldsymbol{x}{'}^{*} = \sqrt{P_M} \boldsymbol{U}^x (\boldsymbol{D}^x)^{-\frac{1}{2}} \boldsymbol{v}^x,
\end{equation}

\subsection{Transmit Amplitude Optimization}

Since $\boldsymbol{\psi}^t$ and $(\boldsymbol{\psi}^t)^H \boldsymbol{R}^t \boldsymbol{\psi}^t$ in (\ref{Pt_OF}) are both real, we have $(\boldsymbol{\psi}^t)^H \boldsymbol{R}^t \boldsymbol{\psi}^t = (\boldsymbol{\psi}^t)^H \text{Re}\left\{\boldsymbol{R}^t\right\} \boldsymbol{\psi}^t$. Similar to problem $(\text{P}_{x})$, $(\text{P}_{t})$ is equivalent to the following problem
\begin{subequations}
	\begin{align}
		\text{P}_{t}^{'}\text{:}\max_{\boldsymbol{\psi}^{t'}}~& (\boldsymbol{\psi}^t{'})^H \boldsymbol{R}^{t}{'} \boldsymbol{\psi}^t{'},\label{Pt'_OF}\\
		s.t.~& (\boldsymbol{\psi}^t{'})^H \boldsymbol{\psi}^t{'} \leq P_M,\\
		& 0 \leq ((\boldsymbol{S}^t)^{-\frac{1}{2}}\boldsymbol{\psi}^t{'})_n \leq 1, \forall n, \label{Pt''_const1}
	\end{align}
\end{subequations}
where $\boldsymbol{\psi}^t{'}=(\boldsymbol{S}^t)^{\frac{1}{2}}\boldsymbol{\psi}^t$, $\boldsymbol{R}^{t}{'}=(\boldsymbol{S}^t)^{-\frac{1}{2}} \text{Re}\left\{\boldsymbol{R}^{t}\right\} (\boldsymbol{S}^t)^{-\frac{1}{2}}$

According to \cite{QCQP}, problem $(\text{P}_{t}^{'})$ is a non-convex problem, which makes it hard to obtain the optimal solution.
Therefore, inspired by~\cite{cf-deriv-1}, we derive the approximated optimal solution to $(\text{P}_{t}^{'})$ in closed-form as follows:
\begin{proposition}\label{prop_3}
	The optimal solution to $(\text{P}_{t}^{'})$ can be approximated by
	\begin{equation}\label{psi_star}
		\boldsymbol{\psi}^{t*} =
		\begin{cases}
			\frac{(\boldsymbol{S}^t)^{-\frac{1}{2}}\boldsymbol{\mu}_{+}}{\left|(\boldsymbol{S}^t)^{-\frac{1}{2}}\boldsymbol{\mu}_{+}\right|} ,~& \left|\boldsymbol{\mu}_{+}\right| \ge \left|\boldsymbol{\mu}_{-}\right|, \\
			\frac{(\boldsymbol{S}^t)^{-\frac{1}{2}}\boldsymbol{\mu}_{-}}{\left|(\boldsymbol{S}^t)^{-\frac{1}{2}}\boldsymbol{\mu}_{-}\right|},~& \left|\boldsymbol{\mu}_{-}\right| > \left|\boldsymbol{\mu}_{+}\right|,
		\end{cases}
	\end{equation}
	where $\left|\cdot\right|$ means taking the Euclidean norm, and 
	\begin{equation}
		\boldsymbol{\mu}_{\pm} = \frac{\text{abs}\left(\text{Re}\left\{\boldsymbol{u}_1\right\}\right) \pm \text{Re}\left\{\boldsymbol{u}_1\right\}}{2}, \label{mu+} 
	\end{equation}
	with $\boldsymbol{u}_1$ being the eigenvector that corresponds to the largest eigenvalue of $\boldsymbol{R}^{t}{'}$, and $\text{abs}\left(\cdot\right)$ meaning taking the absolute value of each element in the vector.
\end{proposition} 
\begin{IEEEproof}
	See Appendix \ref{proof_3}.
\end{IEEEproof}

\subsection{Receive Amplitude Optimization}


Due to the fractional form of \eqref{Pr_OF}, problem $(\text{P}_{r})$ is hard to tackle.
Therefore, we reformulate $(\text{P}_{r})$ as follows \cite{FP_transform}:
\newtheorem{lemma}{\bf Lemma}
\begin{lemma}\label{lemma_2} 
	Problem $(\text{P}_{r})$ is equivalent to 
	\begin{subequations}
		\begin{align}
			\text{P}_{r}^{'}\text{:}\max_{\boldsymbol{\psi}^r, \boldsymbol{\xi}}~& \sum_{l=1}^{L_r} 2 \boldsymbol{\xi}_l^T \boldsymbol{U}_l^T \boldsymbol{\psi}^r - \boldsymbol{\xi}_l^T \left[\boldsymbol{B}_l(\boldsymbol{\psi}^r)\right] \boldsymbol{\xi}_l \label{Pr''_OF}\\
			s.t.~& \boldsymbol{B}_l(\boldsymbol{\psi}^r) = \left[(\boldsymbol{\psi}^r)^T \text{Re}\left\{\boldsymbol{S}^r_l\right\} \boldsymbol{\psi}^r\right] \left[\text{Re}\left\{\boldsymbol{D}_l\right\}\right]^{-1},\\
			& \boldsymbol{\xi}_l \in \mathbb{C}^{N_r\times 1}, 0 \leq \psi_{n}^r \leq 1, 
		\end{align}
	\end{subequations}
	where $\boldsymbol{\xi}=\left\{\boldsymbol{\xi}_1,\cdots,\boldsymbol{\xi}_{N_r}\right\}$ refers to the auxiliary variables, and $\boldsymbol{U}_l \boldsymbol{D}_l \boldsymbol{U}_l^H$ is the spectral decomposition of $\text{Re}\left\{\boldsymbol{R}^r_l\right\}$. 
\end{lemma}
\begin{IEEEproof}
	First, the objective function~\eqref{Pr_OF} of $(\text{P}_{r})$ is equivalent to $\sum_{l=1}^{L_r} (\boldsymbol{U}_l^T \boldsymbol{\psi}^r)^T \left[\boldsymbol{B}_l(\boldsymbol{\psi}^r)\right]^{-1} (\boldsymbol{U}_l^T\boldsymbol{\psi}^r)$.
	Second, the optimal solution of (\ref{Pr''_OF}) is $\boldsymbol{\xi}_l^{*}=\left[\boldsymbol{B}_l(\boldsymbol{\psi}^r)\right]^{-1} (\boldsymbol{U}_l^T\boldsymbol{\psi}^r),l=1,\cdots,N_r$, and the corresponding optimal value equals to $\sum_{l=1}^{L_r} (\boldsymbol{U}_l^T \boldsymbol{\psi}^r)^T \left[\boldsymbol{B}_l(\boldsymbol{\psi}^r)\right]^{-1} (\boldsymbol{U}_l^T\boldsymbol{\psi}^r)$. Therefore, this quadratic transform satisfies the requirements for an equivalent transform as stated in \cite{FP_transform}. 
\end{IEEEproof}

Since the equivalence of problem $(\text{P}_{r})$ and $(\text{P}_{r}^{'})$ is established, we propose the receive beamforming algorithm to solve problem $(\text{P}_{r}^{'})$. Explicitly, $(\text{P}_{r}^{'})$ can be solved by iteratively solving the optimal solutions $\boldsymbol{\xi}^{*}$ and $\boldsymbol{\psi}^{r*}$. 
In the $i$-th iteration, $\boldsymbol{\psi}^{r*}$ and $\boldsymbol{\xi}^{*}$ are solved sequentially, and the value of \eqref{Pr''_OF} $s_i$ is calculated. The iteration halts when $|s_i-s_{i-1}|\le\epsilon'$.
$\boldsymbol{\xi}^{*}$ and $\boldsymbol{\psi}^{r*}$ can be solved based on the following methods:
\subsubsection{Optimization of \texorpdfstring{$\boldsymbol{\xi}$}{}} %
Equation (\ref{Pr''_OF}) in $(\text{P}_{r}^{'})$ is a convex quadratic function of $\boldsymbol{\xi}$ which can be solved by the following equation:
\begin{equation}\label{Pr''_OF_xi}
	\boldsymbol{\xi}_l^{*} = \left[\boldsymbol{B}_l(\boldsymbol{\psi}^r)\right]^{-1} (\boldsymbol{U}_l^T\boldsymbol{\psi}^r),~l=1,\cdots,N_r.
\end{equation}

\subsubsection{Optimization of \texorpdfstring{$\boldsymbol{\psi}^r$}{}} 
The objective function (\ref{Pr''_OF}) in $(\text{P}_{r}^{'})$ can be rewritten as $f(\boldsymbol{\psi}^r) = - (\boldsymbol{\psi}^r)^T \left[  \sum_{l=1}^{L_r} \boldsymbol{\xi}_l^T \left[\text{Re}\left\{\boldsymbol{R}^r_l\right\}\right]^{-1} \boldsymbol{\xi}_l \text{Re}\left\{\boldsymbol{S}^r_l\right\} \right] \boldsymbol{\psi}^r + \left(2 \sum_{l=1}^{L_r} \boldsymbol{\xi}_l^T\boldsymbol{U}_l^T\right) \boldsymbol{\psi}^r$, which is a convex quadratic function of $\boldsymbol{\psi}^r$ with its value constrained in range $[0,1]$. Thus, the optimal solution of $\boldsymbol{\psi}^r$ given $\boldsymbol{\xi}$ can be obtained by the following proposition: 
\begin{proposition}
	\label{prop_4}
	The optimal solution $\boldsymbol{\psi}^{r*} = \left[\boldsymbol{\psi}^{r*}_1,\cdots,\boldsymbol{\psi}^{r*}_{N_r}\right]$ of problem $(\text{P}_{r}^{'})$ can be given by:
	\begin{equation}\label{Pr''_OF_psi}
		\boldsymbol{\psi}^{r*}_n = 
		\begin{cases}
			\boldsymbol{\psi}^{*}_n, & \text{if } 0 \le \boldsymbol{\psi}^{*}_n \le 1 \\
			0, & \text{if } \boldsymbol{\psi}^{*}_n < 0 \\
			1, & \text{if } \boldsymbol{\psi}^{*}_n > 1
		\end{cases},
		n=1,\cdots,N_r,
	\end{equation}
	where $\boldsymbol{\psi}^{*} = \left[\boldsymbol{\psi}^{*}_1,\cdots,\boldsymbol{\psi}^{*}_{N_r}\right] = \left(2 \sum_{l=1}^{L_r} \boldsymbol{U}_l \boldsymbol{\xi}_l\right) \cdot \left[  \sum_{l=1}^{L_r}  \boldsymbol{\xi}_l^T \left[\text{Re}\left\{\boldsymbol{R}^r_l\right\}\right]^{-1} \boldsymbol{\xi}_l \text{Re}\left\{\boldsymbol{S}^r_l\right\} \right]^{-1} $ is the global maximum point of function $f(\boldsymbol{\psi}^r)$. 
\end{proposition}
\begin{IEEEproof}
	Since $- \left[  \sum_{l=1}^{L_r}  \boldsymbol{\xi}_l^T \left[\text{Re}\left\{\boldsymbol{R}^r_l\right\}\right]^{-1} \boldsymbol{\xi}_l \text{Re}\left\{\boldsymbol{S}^r_l\right\} \right]$ is negative semi-definite, $f(\boldsymbol{\psi}^r)$ is a concave function. According to the property of concave functions, $f(\boldsymbol{\psi}^{*}_n) \ge f(\boldsymbol{\psi}^{r*}_n) \ge f(\boldsymbol{\psi}^{r}_n), \forall 0 \le \boldsymbol{\psi}^{r}_n \le 1$ if $\boldsymbol{\psi}^{*}_n < 0$. And the proof of the case where $\boldsymbol{\psi}^{*}_n > 1$ can be established similarly. 
\end{IEEEproof}

\section{Simulation Results}
In this section, simulation results are provided to evaluate the detection performance of the proposed RHS-enabled radar system. The radar signal frequency is set as $30$ GHz. The elements spacings of the RHS and the phased array are $\lambda/3$ \cite{ele_spac} and $\lambda/2$, respectively. We set the waveguide refractive index $\nu$ as $\sqrt{3}$, and attenuation factor $\alpha$ as $5$. 
The space of interest is $\theta\in[0,2\pi),~\varphi\in\left\{0,\pi/10\right\}$, and the azimuthal region $[0,2\pi)$ is evenly partitioned into $4$ angular grids.
We suppose there are two targets respectively located in directions $(\pi/4,\pi/6)$ and $(3\pi/4,\pi/6)$, which corresponds to the hypothesis $\mathcal{H}_j$. 

\begin{figure*}[!t]
	\centering
	\subfloat[]{
		\includegraphics[height=1.5in]{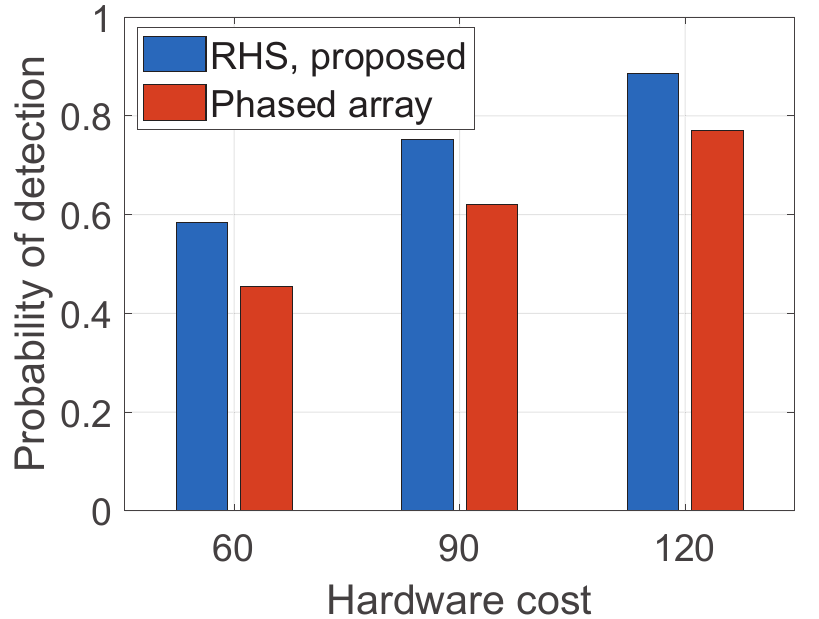}
	}
	\hspace{0.15in} 
	\
	\subfloat[]{
		\includegraphics[height=1.5in]{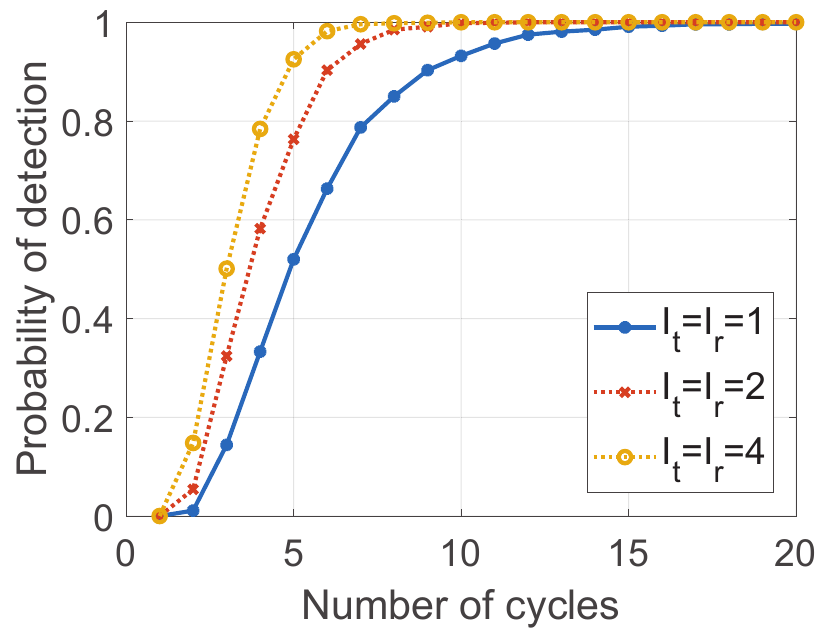}
	}
	\hspace{0.15in}
	\subfloat[]{
		\includegraphics[height=1.5in]{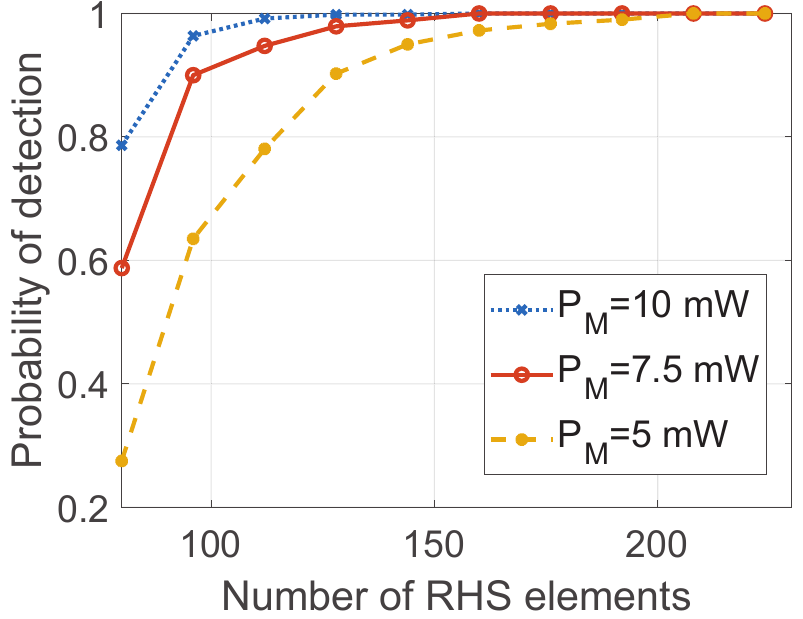}
	}
	\caption{(a) Probability of detection v.s. hardware cost; (b) probability of detection v.s. number of cycles; (c) probability of detection v.s. number of RHS elements.} 
	\label{f_s}
\end{figure*}

Fig. 2(a) depicts the probability of detection versus hardware cost\footnote{The number of antenna element of the RHS and the phased array are set as $(60,10),(90,15),(120,20)$, respectively.} obtained by the proposed RHS scheme and the phased array scheme given the same number of cycles $6$.
Assume that the hardware cost of a phased array antenna is $\tau$ times of that of an RHS element on average. According to~\cite{power_effi}, $\tau$ is set as $6$. For simplicity, the cost of an RHS element is normalized to $1$. In the phased array scheme, considering the transmission line loss, we set the amplitude attenuation of the signal from the antenna to the feed as $0.9$~\cite{PA_attenu}. 
The optimization technique of the phased array is designed based on the method in~\cite{PA_algor}. 
We observe that the probability of detection of both schemes increase as the hardware cost increases, implying that the detection accuracy can be improved at the cost of increasing hardware expenses. Moreover, it can be seen that given the same cost, the detection probability of the proposed RHS scheme has an increase of about $0.13$ compared to that of the phased array one, indicating that the RHS can achieve a higher detection accuracy compared to the phased array.

Fig. 2(b) illustrates the probability of detection versus different numbers of cycles obtained by the proposed RHS-enabled scheme with different lengths of the radar signals. The numbers of RHS elements are set as $N_t=N_r=60$. 
It can be seen that the probability of detection increases as the number of cycles increases, verifying that the detection accuracy can be improved by performing more cycles.
In addition, given the same number of cycles, the detection probability of the RHS-enabled radar increases as the length of the radar signal increases, indicating that the detection accuracy can be improved by transmitting longer radar signals.

Fig. 2(c) depicts the probability of detection versus the number of RHS elements with different values of $P_M$. 
The noise variance is set as $\sigma^2=1$. And we set the number of cycles as $3$. 
As the radiation power increases, the detection probability increases given the element number.
Besides, it can be observed that the probability of detection increases as the number of RHS elements increases, implying that the detection performance can be improved by increasing the physical size of the RHS.

\section{Conclusion}
In this paper, we have developed an RHS-enabled radar system for multi-target detection, where two RHSs are co-located to serve as transmit and receive antennas. To optimize the detection accuracy, we have formulated a relative entropy based optimization problem 
and proposed an iterative algorithm to jointly optimizing the radar waveform and the beamforming vectors of the transmit and receive RHSs. Simulation results show that: 1) the proposed RHS-enabled radar can increase the probability of detection by $0.13$ on average compared to phased array radars when six cycles are performed given the same hardware cost; 2) The detection accuracy of the proposed RHS-enabled radar system can be improved by increasing the physical aperture of the RHS, the radiation power, and the length of the radar signal. 

\appendices

\section{Proof of Proposition~\ref{prop_1}}\label{proof_1}
The relative entropy between distributions of hypotheses $\mathcal{H}_i$ and $\mathcal{H}_j~(i\neq j)$ can be given by
\begin{equation}
	\begin{aligned}
		& D\left(p(\boldsymbol{y} | \mathcal{H}_i) \Vert p(\boldsymbol{y} | \mathcal{H}_j)\right) \\
		=& -\frac{1}{2}\left[ \left(\boldsymbol{u}_j-\boldsymbol{u}_i\right)^H \boldsymbol{\Sigma}^{-1} \boldsymbol{u}_i + \boldsymbol{u}_i^H\boldsymbol{\Sigma}^{-1}\left( \boldsymbol{u}_j-\boldsymbol{u}_i \right) \right.\\
		& \qquad\qquad \left. + \boldsymbol{u}_i^H\boldsymbol{\Sigma}^{-1}\boldsymbol{u}_i - \boldsymbol{u}_j^H\boldsymbol{\Sigma}^{-1}\boldsymbol{u}_j \right].
	\end{aligned}
\end{equation}
Therefore, the distance between $\mathcal{H}_i$ and $\mathcal{H}_j~(i\neq j)$ can be expressed as
\begin{equation}
	\begin{aligned}
		d(\mathcal{H}_i,\mathcal{H}_j|\boldsymbol{y}) &= 
		D\left(p(\boldsymbol{y} | \mathcal{H}_i) \Vert p(\boldsymbol{y} | \mathcal{H}_j)\right) 
		+D\left(p(\boldsymbol{y} | \mathcal{H}_j) \Vert p(\boldsymbol{y} | \mathcal{H}_i)\right)\\
		&= \left(\boldsymbol{u}_i-\boldsymbol{u}_j\right)^H \boldsymbol{\Sigma}^{-1} \left(\boldsymbol{u}_i-\boldsymbol{u}_j\right).
	\end{aligned}
\end{equation}

\section{Proof of Proposition~\ref{prop_3}}\label{proof_3}
According to spectral theorem, $\boldsymbol{R}^t{'}$ can be approximated by 
$\gamma_1 \boldsymbol{u}_1 \boldsymbol{u}_1^T$, 
where $\gamma_1$ is its largest eigenvalue and $\boldsymbol{u}_1$ is the corresponding eigenvector.
Therefore, $(\text{P}_{t}^{'})$ can be approximated by maximizing 
$\gamma_1 \left|\boldsymbol{u}_1^T \boldsymbol{\psi}^{t}{'}\right|^2$ 
subject to $\left|\boldsymbol{\psi}^{t}{'}\right|^2 \le P_M$.

Since $\boldsymbol{\psi}_n^{t}{'} \ge 0, \forall n$, we have 
\begin{equation}
	\left|\boldsymbol{u}_1^T \boldsymbol{\psi}^{t}{'}\right|^2 \le \max \left\{ \max\left\{\left|\boldsymbol{\mu}_{+}^T \boldsymbol{\psi}^{t}{'}\right|^2\right\}, \max\left\{\left|\boldsymbol{\mu}_{-}^T \boldsymbol{\psi}^{t}{'}\right|^2\right\} \right\},
\end{equation}
where $\boldsymbol{\mu}_{+}$ and $\boldsymbol{\mu}_{-}$ are defined in \eqref{mu+}. 

Based on Cauchy–Schwarz inequality, we have
\begin{equation}
	\left|\boldsymbol{\mu}_{+}^T \boldsymbol{\psi}^{t}{'}\right|^2 \le \left|\boldsymbol{\mu}_{+}^T\right|^2 \left| \boldsymbol{\psi}^{t}{'}\right|^2 \le P_M \left|\boldsymbol{\mu}_{+}\right|^2.
\end{equation}
The two sides are equal if and only if $\boldsymbol{\psi}^{t}{'}=C_{+}\boldsymbol{\mu}_{+}$ with $C_{+}$ being a constant.
Similarly, $\left|\boldsymbol{\mu}_{-}^T \boldsymbol{\psi}^{t}{'}\right|^2$ is maximized if and only if $\boldsymbol{\psi}^{t}{'}=C_{-}\boldsymbol{\mu}_{-}$ with $C_{-}$ being a constant.
Besides, since $\boldsymbol{\psi}^{t*}$ needs to satisfy (\ref{Pt_const1}), let $C_{+}=1/|(\boldsymbol{S}^t)^{-\frac{1}{2}}\boldsymbol{\mu}_{+}|$ and $C_{-}=1/|(\boldsymbol{S}^t)^{-\frac{1}{2}}\boldsymbol{\mu}_{-}|$. We then obtain \eqref{psi_star} in Proposition \ref{prop_3}. 

\end{document}